\documentclass[acmtog]{acmart}

\usepackage{algorithmicx, acmart-taps}
\usepackage{algorithm}
\usepackage{algpseudocode}
\usepackage{graphicx}
\usepackage{xcolor}
\usepackage{subcaption}

\usepackage{colortbl}    %
\usepackage{pdfpages}

\definecolor{linkblue}{RGB}{0,92,175}

\definecolor{changeColor}{rgb}{0,0.0,0}

\def\algorithmicinput{\textbf{Input:}}
\def\algorithmicoutput{\textbf{Output:}}
\def\Input{\item[\algorithmicinput]}
\def\Output{\item[\algorithmicoutput]}

\copyrightyear{2026}
\acmYear{2026}
\setcopyright{cc}
\setcctype{by}
\acmConference[SIGGRAPH Conference Papers '26]{Special Interest Group on Computer Graphics and Interactive Techniques Conference Conference Papers}{July 19--23, 2026}{Los Angeles, CA, USA}
\acmBooktitle{Special Interest Group on Computer Graphics and Interactive Techniques Conference Conference Papers (SIGGRAPH Conference Papers '26), July 19--23, 2026, Los Angeles, CA, USA}
\acmDOI{10.1145/3799902.3811189}
\acmISBN{979-8-4007-2554-8/2026/07}

\acmSubmissionID{1130}

\begin{document}

\title{Fast Sparse Matrix Permutation for Mesh-Based Direct Solvers}

\author{Behrooz Zarebavani}
\affiliation{%
  \department{Department of Computer Science}
  \institution{University of Toronto}
  \streetaddress{6 King's College Road}
  \city{Toronto}
  \state{ON}
  \postcode{M5S 3H5}
  \country{Canada}}
\email{behrooz.zarebavani@gmail.com}
\orcid{https://orcid.org/0000-0003-2060-9596}
\authornote{Joint first author}

\author{Ahmed H. Mahmoud}
\affiliation{%
  \department{Computer Science \& Artificial Intelligence Laboratory}
  \institution{Massachusetts Institute of Technology}
  \streetaddress{32 Vassar St}
  \city{Cambridge}
  \state{MA}
  \postcode{02139}
  \country{USA}}
\email{ahdhn@mit.edu}
\orcid{0000-0003-1857-913X}
\authornotemark[1]

\author{Ana Dodik}
\affiliation{%
  \department{Computer Science \& Artificial Intelligence Laboratory}
  \institution{Massachusetts Institute of Technology}
  \streetaddress{32 Vassar St}
  \city{Cambridge}
  \state{MA}
  \postcode{02139}
  \country{USA}}
\email{anadodik@mit.edu}
\orcid{https://orcid.org/0000-0003-4391-8877}

\author{Changcheng Yuan}
\affiliation{%
  \department{Department of Computer Science and Engineering}
  \institution{Texas A\&M University}
  \streetaddress{435 Nagle St}
  \city{College Station}
  \state{TX}
  \postcode{77843}
  \country{USA}}
\email{eric.yuan.cc@gmail.com}
\orcid{https://orcid.org/0009-0005-9901-3454}

\author{Serban D. Porumbescu}
\affiliation{%
  \department{Department of Electrical and Computer Engineering}
  \institution{University of California, Davis}
  \streetaddress{One Shields Avenue}
  \city{Davis}
  \state{CA}
  \postcode{95616}
  \country{USA}}
\email{sdporumbescu@ucdavis.edu}
\orcid{0000-0003-1523-9199}

\author{John D. Owens}
\affiliation{%
  \department{Department of Electrical and Computer Engineering}
  \institution{University of California, Davis}
  \streetaddress{One Shields Avenue}
  \city{Davis}
  \state{CA}
  \postcode{95616}
  \country{USA}}
\email{jowens@ece.ucdavis.edu}
\orcid{0000-0001-6582-8237}

\author{Maryam Mehri Dehnavi}
\affiliation{%
  \department{Department of Computer Science}
  \institution{University of Toronto}
  \streetaddress{6 King's College Road}
  \city{Toronto}
  \state{ON}
  \postcode{M5S 3H5}
  \country{Canada}}
\email{mmehride@cs.toronto.edu}
\orcid{https://orcid.org/0000-0002-2719-8788}
\affiliation{%
  \institution{NVIDIA Research}
  \country{USA}
}
\email{mdehnavi@nvidia.com}
\authornote{Joint last author}

\author{Justin Solomon}
\affiliation{%
  \department{Computer Science \& Artificial Intelligence Laboratory}
  \institution{Massachusetts Institute of Technology}
  \streetaddress{32 Vassar St}
  \city{Cambridge}
  \state{MA}
  \postcode{02139}
  \country{USA}
}
\email{jsolomon@mit.edu}
\orcid{0000-0002-7701-7586}
\authornotemark[2]

\renewcommand{\shortauthors}{Zarebavani and Mahmoud et al.}




\begin{abstract}
    We present a fast sparse matrix permutation algorithm tailored to linear systems arising from triangle meshes. Our approach produces nested-dissection-style permutations while significantly reducing permutation runtime overhead. Rather than enforcing strict balance and separator optimality, the algorithm deliberately relaxes these design decisions to favor fast partitioning and efficient elimination-tree construction. Our method decomposes permutation into patch-level local orderings and a compact quotient-graph ordering of separators, preserving the essential structure required by sparse Cholesky factorization while avoiding its most expensive components. We integrate our algorithm into vendor-maintained sparse Cholesky solvers on both CPUs and GPUs. Across a range of graphics applications, including single factorizations and repeated factorizations, our method reduces permutation time and improves the sparse Cholesky solve performance by up to $6.27\times$. Our code is available at \href{https://github.com/BehroozZare/fast-permute}{\textcolor{linkblue}{https://github.com/BehroozZare/fast-permute}}. %
\end{abstract}


\maketitle




\section{Introduction}
\label{sec:intro}

Sparse linear solvers are central computational kernels in computer graphics. Many mesh-centric workloads (e.g., parameterization, deformation, physical simulation, and geometric flows) solve large positive semidefinite systems, for which sparse Cholesky factorization is often the most practical and convenient choice due to its robustness. Nevertheless, in many graphics pipelines, overall runtime remains dominated by the cost of the linear solver.

\begin{figure}[h]
  \centering
  \includegraphics[width=0.99\linewidth]{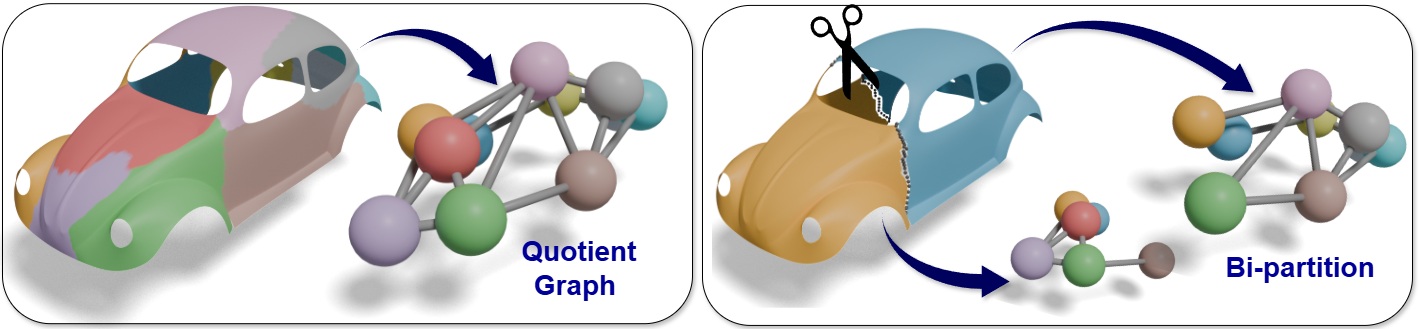}
\caption{\textbf{Scalable fill-reducing permutation computation on a triangular mesh.} Our patch-based, nested-dissection-style method compresses the triangular mesh domain into a small quotient graph and reuses this compression across recursive calls, improving scalability over state-of-the-art methods. We achieve a 6.62$\times$ speedup over the highly optimized NVIDIA cuDSS by accelerating its permutation computation.}
  \label{fig:teaser}
\end{figure}

\noindent\begin{minipage}{\linewidth}
    \centering
    \includegraphics[width=0.8\linewidth]{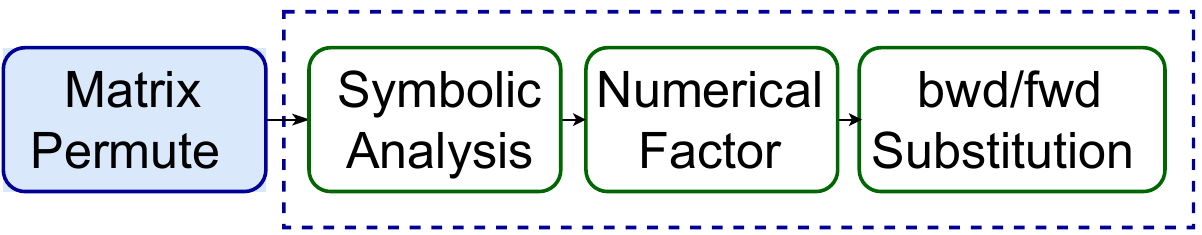}

    {\small\textbf{Sparse Cholesky factorization pipeline.}}
\end{minipage}

\begin{figure}
    \centering
    \includegraphics[width=0.98\linewidth]{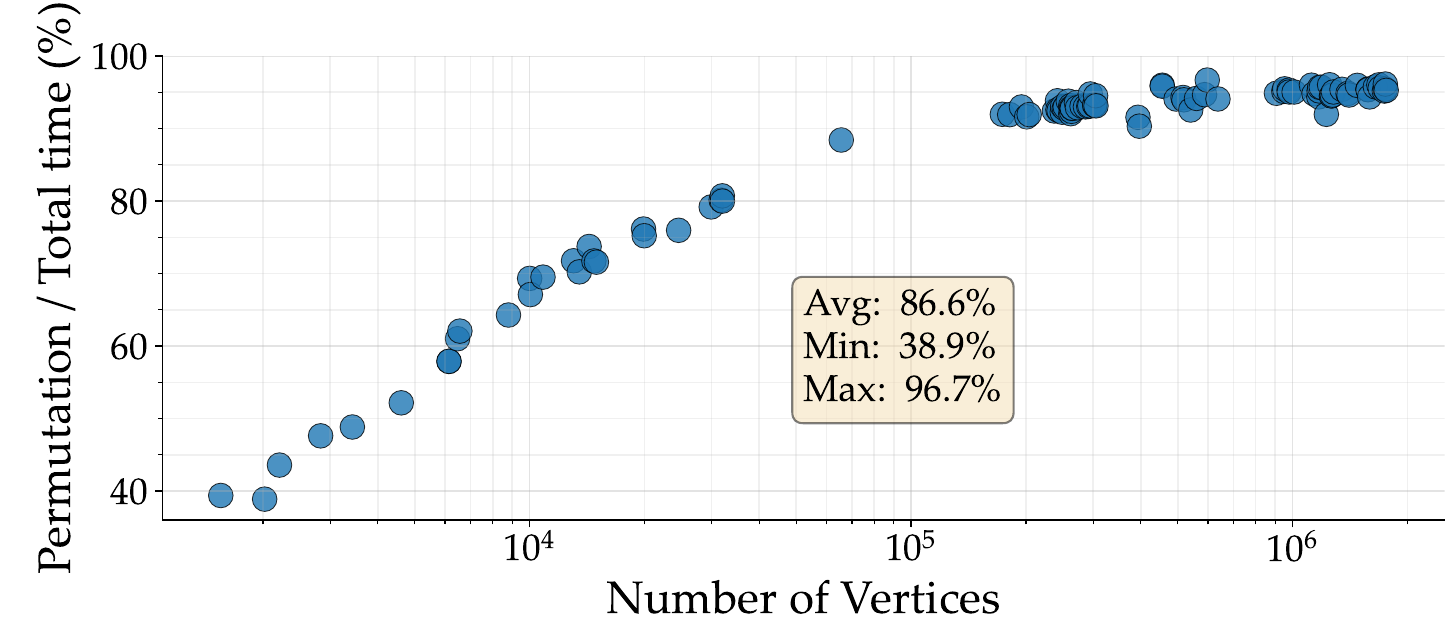}
    \caption{Fraction of time spent on matrix permutation to end-to-end time for solving mean curvature flow~\cite{Desbrun:1999:IFO} for different inputs.}
    \label{fig:perm_ratio}
\end{figure}

Sparse Cholesky solvers are often accelerated by amortizing an upfront symbolic analysis phase that inspects the sparsity pattern and prepares data structures for efficient factorization and forward/backward solves. While recent solvers like Apple Accelerate~\cite{Apple:2025:Accelerate}, Intel MKL~\cite{Intel:2025:MKL}, and NVIDIA cuDSS~\cite{NVIDIA:2025:cudss} significantly improve the symbolic and numerical stages, the symbolic phase still contains a key scalability bottleneck, i.e., computing a \emph{fill-reducing permutation}. Permutation is an essential preprocessing step in the symbolic phase; it reorders the rows and columns of the matrix to make the Cholesky factors as sparse as possible. In our experiments (see Figure~\ref{fig:perm_ratio}), permutation accounts for an average of 86\% of the total runtime, reaching 96\% for large meshes in the newly-introduced, highly-optimized NVIDIA cuDSS solver.

Finding the optimal fill-reducing permutation is NP-complete~\cite{Yannakakis:1981:CTM}, so practical solvers rely on heuristics~\cite{Davis:2006:DMF}. High-performance solvers predominantly use nested dissection (e.g., METIS~\cite{Karypis:2013:MSG}) or minimum-degree orderings (e.g., AMD~\cite{Amestoy:1996:AAM}).
Nested dissection typically yields higher-quality orderings and enables parallel factorization via its hierarchical structure, while AMD has low ordering cost but can produce higher fill-in on mesh-like problems~\cite{Lipton:1979:GND}.

In current practice, permutation is treated as a standalone preprocessing step and applied as a black box before symbolic analysis. This separation leads to loss of nested-dissection hierarchical information, which can be reused in permutation and symbolic analysis. For example, NVIDIA cuDSS requires this structure in the form of an \emph{elimination tree} for its symbolic analysis. This separation creates a gap between high-quality orderings that also provide reusable hierarchy information for symbolic analysis, and faster orderings that focus on permutation alone. The latter not only tend to increase fill-in, but also force solvers to recompute symbolic structures that could otherwise be reused from a nested-dissection hierarchy.

In this paper, we introduce a sparse matrix permutation algorithm tailored to mesh-derived linear systems. Our method produces nested-dissection-style orderings that exploit mesh structure while remaining efficient to compute. Crucially, it preserves the reusable separator hierarchy needed for symbolic analysis, while achieving a better quality--runtime trade-off through faster separator computation and efficient integration of minimum-degree techniques such as AMD. Compared to state-of-the-art permutation tools, our algorithm achieves up to 10.27$\times$ speedup (4.58$\times$ geometric mean) in ordering time and delivers up to 6.62$\times$ end-to-end speedup (3.51$\times$ geometric mean) when solving linear systems.

Our key insight is that nested-dissection permutation is often dominated by the effort spent enforcing strict balance and minimizing separator size, as done by general-purpose partitioners such as METIS~\cite{Karypis:2013:MSG} and PT-Scotch~\cite{Chevalier:2008:PAT}. For mesh-derived systems, relaxing these constraints dramatically reduces permutation time; although this produces lower-quality separators, the reduced preprocessing overhead outweighs the increased fill-in in practice. This motivates a multi-stage permutation strategy that decouples ordering speed from ordering quality while retaining reusable symbolic structure. In particular, our method constructs the \emph{elimination tree} alongside the permutation, enabling direct reuse for symbolic scheduling and jointly accelerating both permutation and symbolic analysis. Our code is available at \href{https://github.com/BehroozZare/fast-permute}{\textcolor{linkblue}{https://github.com/BehroozZare/fast-permute}}.



\begin{figure*}
    \centering
    \includegraphics[width=0.9\textwidth]{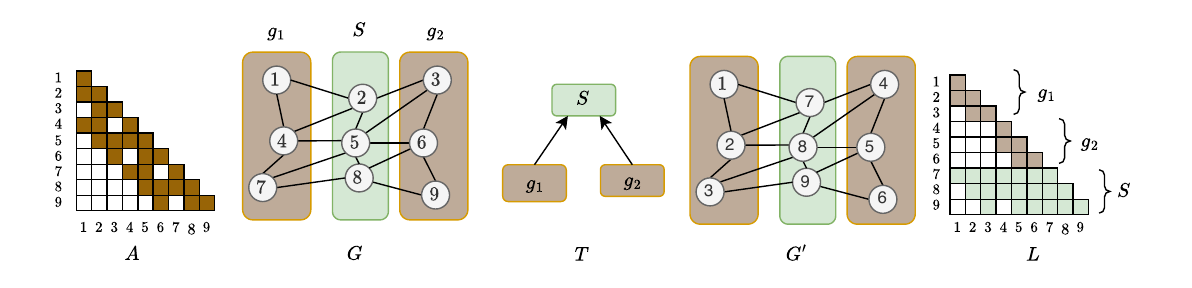}
    \caption{\textbf{Illustration of a single nested-dissection step.} From left to right: the input matrix $A$, its graph $G$, the corresponding elimination tree $T$, the permuted graph $G'$, and the sparsity pattern of the Cholesky factor $L$. The separator vertices $S$ are ordered after the vertices in $g_1$ and $g_2$. The elimination tree encodes the dependencies induced by this order, i.e., computations for $g_1$ and $g_2$ are independent, while computations for $S$ can proceed only after both have completed.}
    \label{fig:background:SepExp}
\end{figure*}

\begin{figure}[h]
    \centering
    \includegraphics[width=0.9\linewidth]{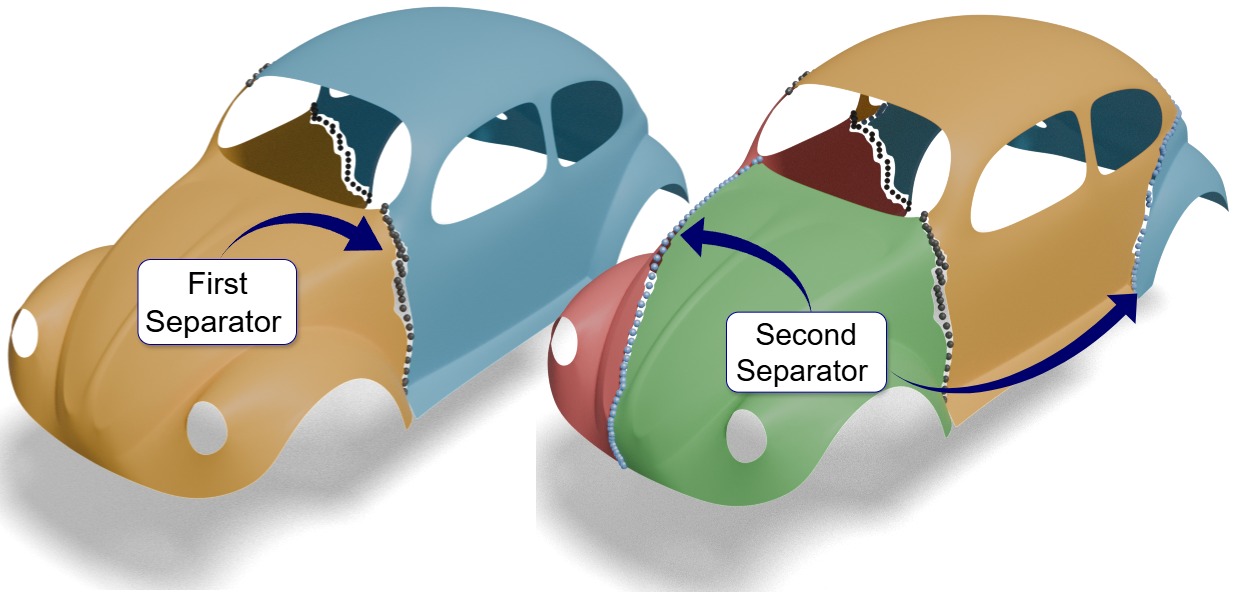}
    \caption{\textbf{Separator set on a mesh.} Nested dissection recursively partitions the mesh by choosing small sets of vertices (shown as green points) whose removal separates the remaining mesh into disconnected subdomains. The first separator splits the mesh into two large regions  (left). The second separator further partitions one of these regions (right).}
    \label{fig:sep}
\end{figure}
\section{Background}
\label{sec:background}
Given a sparse symmetric matrix $A \in \mathbb{R}^{n \times n}$, we define an undirected graph
$G = (V_G, E_G)$ where each vertex $i \in V_G$ corresponds to row/column $i$ of $A$, and there is
an edge $(i,j) \in E_G$ if and only if $A_{ij} \neq 0$. We refer to $G$ as the \emph{graph} of $A$.

Nested dissection~\cite{Lipton:1979:GND} recursively partitions $G$ by finding a small set of
vertices $S$ (i.e., \emph{separators}) whose removal separates $G$ into two subgraphs, $g_1$ and $g_2$ (see Figure~\ref{fig:sep}). The resulting permutation orders the vertices in $g_1$ and $g_2$ before the vertices in $S$. During Cholesky factorization, this means that the unknowns associated
with $g_1$ and $g_2$ are eliminated first and the separator unknowns are eliminated later~\cite{Scott:2023:AFS}.
The dependencies induced by this elimination order are captured by the \emph{elimination tree}
(etree): its nodes represent elimination groups (i.e., blocks of rows/columns) and its
edges indicate that eliminating one group produces updates to another. For a single dissection,
$g_1$ and $g_2$ form two independent subproblems that meet at the separator $S$ (the
parent). Thus, fill-in is created only within $g_1$, within $g_2$, and within the separator
block, there is no fill-in between $g_1$ and $g_2$ (see Figure~\ref{fig:background:SepExp}). Nested dissection
repeats this process recursively on the subgraphs until reaching a prescribed stopping criterion
(e.g., recursion depth).

Many sparse linear systems in geometry processing arise from FEM discretizations. For such
operators (e.g., Laplace-Beltrami), there is a one-to-one correspondence
between mesh vertices and unknowns (rows/columns) in the sparse linear system, and thus between mesh vertices and the vertices of $G$. In this
setting, $G$ matches the mesh adjacency graph. %
 We exploit this correspondence to move between the mesh and the matrix graph.%

Finally, we define a \emph{patch} as a connected subset of mesh vertices $V_M$. Formally, a patch $P \subseteq V_M$
is a set of vertices whose induced submesh is connected. A collection of patches $\{P_k\}$ forms a
partition of the mesh if $\bigcup_k P_k = V_M$ and $P_k \cap P_\ell = \emptyset$ for $k \neq \ell$. Each patch induces a subgraph of $G$ through $V_M \leftrightarrow V_G$
mapping. The \emph{quotient graph} is the graph whose nodes correspond to patches and whose edges
capture patch adjacency induced by edges in $G$.



\section{Related Work}
\label{sec:related}

Finding a permutation that minimizes fill-in is NP-complete~\cite{Ashkiani:2018:ADH}. Exact methods based on nonserial dynamic programming~\cite{Bertele:1969:NDP, Bertele:1972:NDP} do not scale and are not used in practice. Practical solvers rely on heuristics. \citet{Bichot:2013:GP} and \citet{Schulz:2018:GP} give an overview of fill-reducing orderings. We review widely-used matrix permutation techniques followed by prior work that leverages nested dissection in symbolic analysis.

\paragraph{Minimum-degree orderings.}
\label{sec:related:md}
Minimum-degree (MD) methods construct an elimination order by repeatedly selecting a vertex with (approximately) minimum degree in the graph, aiming to introduce the fewest fill edges~\cite{Tinney:1967:DSS}. Each elimination turns the vertex's neighborhood into a clique, so degrees must be updated after every step; these updates dominate runtime and motivate fast, inexact update heuristics~\cite{George:1973:NDO, George:1980:FIM, Amestoy:1996:AAM}. Approximate Minimum Degree (AMD)~\cite{Amestoy:1996:AAM} is the standard implementation in practice and recent work reduces its cost via improved update strategies~\cite{Cummings:2021:FMD} and GPU implementations~\cite{Chang:2025:Chang}. MD methods are fast, but their local, greedy decisions can yield less balanced elimination trees and expose less parallelism than nested-dissection.

\paragraph{Nested dissection.}
\label{sec:related:nd}
Nested dissection (ND) reduces fill by recursively partitioning the matrix graph using small vertex separators and ordering separator vertices after interior vertices~\cite{George:1973:NDO}. This divide-and-conquer structure typically achieves strong fill reduction on PDE/mesh graphs and yields more parallelism due to balanced recursion~\cite{Lipton:1979:GND}. For completeness, we include a brief didactic background on how permutation affects fill-in and solver parallelism with examples in Supplemental Materials A.
Modern ND implementations rely on multilevel graph partitioners that coarsen, partition, and refine the graph to compute separators efficiently. Once subgraphs become small, they often switch to simple local orderings (e.g., breadth-first variants~\cite{Cuthill:1969:RTB}). Widely used tools include METIS~\cite{Karypis:1998:AFA, Karypis:2013:MSG}, ParMETIS~\cite{Karypis:2013:PPG}, and PT-Scotch~\cite{Chevalier:2008:PAT}.

We build on ND's recursive structure, but observe that top levels of the recursion can dominate ordering time. Since separator computation is NP-complete, reducing the problem size can substantially accelerate this step. Accordingly, we compress the mesh into patches, reuse this patching across recursive calls, and introduce a relaxed variant improving the quality--runtime trade-off in practice.

\paragraph{Exploiting the ND hierarchy in symbolic analysis.}
\label{sec:related:graphicSpecific}
Beyond producing a permutation, ND induces a hierarchy that can be reused in symbolic analysis and downstream factorization. \citet{Herholz:2018:FO} and \citet{Herholz:2020:SCU} accelerate updates by identifying affected regions in the elimination tree and restricting refactorization and triangular solves to those regions. NASOQ~\cite{Cheshmi:2020:NASOQ} leverages a related tree structure in constraint-based quadratic programming. Parth~\cite{Zarebavani:2025:Parth} reuses the elimination tree to accelerate reordering when the sparsity pattern changes over time. Collectively, these works highlight that explicitly creating and exposing the ND-induced hierarchy enables substantial reuse beyond a one-shot ordering.



\section{Our Algorithm}
\label{sec:method}
\begin{figure*}
    \centering
    \begin{subfigure}[t]{0.33\textwidth}
        \centering
        \includegraphics[width=\textwidth]{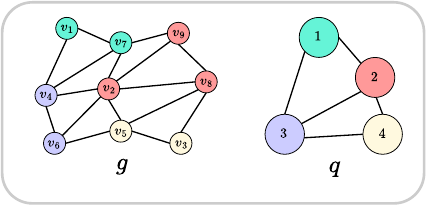}
        \caption{Step 1: Compute quotient graph}
        \label{fig:sepcomp:quotient}
    \end{subfigure}\hfill\begin{subfigure}[t]{0.33\textwidth}
        \centering
        \includegraphics[width=\textwidth]{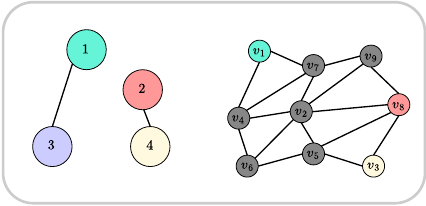}
        \caption{Step 2 \& 3: Bi-partition \& separator superset.}
        \label{fig:sepcomp:refine}
    \end{subfigure}\hfill\begin{subfigure}[t]{0.33\textwidth}
        \centering
        \includegraphics[width=\textwidth]{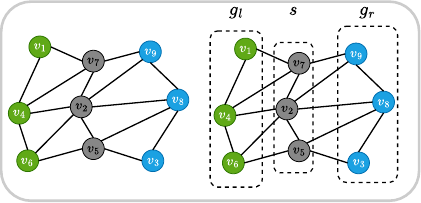}
        \caption{Step 3 \& 4: Refine \& compute sub-graphs}
        \label{fig:sepcomp:subgraphs}
    \end{subfigure}

    \caption{\textbf{Separator computation.} For the current subgraph $\textit{g}$, we form a quotient graph $\textit{q}$, bipartition $\textit{q}$, lift the partition back to $V_\textit{g}$ to obtain a separator superset, refine it to reduce its size while maintaining balance, and then extract the left and right subgraphs for the next recursion level.}
    \label{fig:Step3:SeparatorComp}
\end{figure*}
\paragraph{Overview.}
Our algorithm accelerates fill-reducing nested-dissection permutation by restricting separator computation to a coarse search space, i.e., the quotient graph. In the limiting case where each patch represents a single mesh vertex, our algorithm reduces to classical nested-dissection~\cite{George:1973:NDO}. However, since patches contain multiple vertices, separator decisions are made at the patch level, trading some ordering quality for a substantial reduction in permutation cost, since the size of quotient graph is significantly smaller than the input graph $G$. Concretely, we (i)~compute a patch partition of the mesh on the GPU, (ii)~lift this partition to the matrix graph to form a \emph{group map} ($gmap$) that clusters graph vertices by patch, (iii)~build an elimination tree (\textit{etree}) using nested dissection guided by these groups, and (iv)~assemble a global permutation by traversing the tree in nested-dissection order (see Algorithm~\ref{alg:AlgNameSteps}). We use the GPU only for the first step (computing patches). The rest of the algorithm runs on the CPU.

Our design has multiple advantages. First, patches reduce the separator search space by allowing nested dissection to operate on the quotient graph rather than the full matrix graph. Since patch construction is inexpensive and highly parallel for meshes, as shown in prior work (e.g., RXMesh~\cite{Mahmoud:2021:RAG}, MeshTaichi~\cite{Chang:2022:MAC}), this reduction dramatically decreases permutation runtime. Second, unlike classical nested-dissection implementations (e.g., METIS), which repeatedly coarsen the graph at each recursion level, our method constructs a single quotient graph and reuses it throughout the recursion. This reuse avoids redundant coarsening work and further accelerates permutation. Finally, our algorithm generates the \textit{etree} as part of the permutation process, eliminating a costly step in the symbolic Cholesky phase. Together, our algorithm reduces preprocessing overhead and translates directly into faster end-to-end sparse direct solves.

\paragraph{Inputs and outputs.}
Our algorithm inputs a sparse symmetric positive semidefinite system matrix $A$, a mesh $M$, and a nested-dissection depth parameter \textit{nd\_level}. It outputs a fill-reducing permutation vector $\textit{perm}$ and the corresponding $\textit{etree}$. We use the mesh only to generate patches and the associated quotient graph. Thus, if the patch partition is provided, the mesh input is unnecessary.

\begin{algorithm}[tp]
    \begin{algorithmic}[1]
        \Input $A$, $M$, \textit{nd\_level}
        \Output \textit{perm}, \textit{etree}
        \Statex \textcolor{gray}{/* Step 1: Obtain a patch partition */}
        \State $\textit{patches} \leftarrow \textsc{GetPatches}(M)$ \Comment{or user-provided}
        \Statex \textcolor{gray}{/* Step 2: Build the graph and lift patches to graph groups */}
        \State $\textit{G},\, \textit{gmap} \leftarrow \textsc{BuildGraphAndGroups}(A, M, \textit{patches})$
        \Statex \textcolor{gray}{/* Step 3: Construct etree via patch-guided nested dissection */}
        \State $\textit{etree} \leftarrow \textsc{BuildEtree}(\textit{G}, \textit{gmap}, \textit{nd\_level})$
        \Statex \textcolor{gray}{/* Step 4: Assemble a global permutation from the tree */}
        \State $\textit{perm} \leftarrow \textsc{ComputePerm}(\textit{etree}, \textit{G})$
    \end{algorithmic}
    \caption{Algorithm overview.}
    \label{alg:AlgNameSteps}
\end{algorithm}

\begin{figure}
    \centering
    \includegraphics[width=0.8\linewidth]{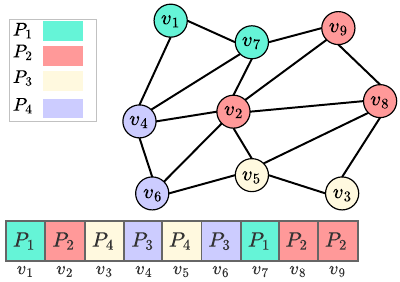}
    \caption{\textbf{Mapping transformation.} The patching algorithm assigns mesh vertices to patches ($P_1 \dots P_4$), from which we construct the group map (\textit{gmap}).}
    \label{fig:Step1:Mapping}
\end{figure}

\subsection{Patching and Constructing the Group Map}
Our algorithm (Algorithm~\ref{alg:AlgNameSteps}) starts with a preprocessing stage. It includes generating mesh patches and constructing the group map, providing a coarse structure that reduces the separator search space.

\paragraph{Step 1: Patch computation}
Many graphics applications compute patch-like partitions (e.g., for remeshing, clustering, or domain decomposition) which we can use directly. Otherwise, we compute \textit{patches} using a fast parallel routine. Our implementation uses the GPU patching method from RXMesh~\cite{Mahmoud:2021:RAG}.

\paragraph{Step 2: Lift patches to the matrix graph.} We construct the graph of $A$ (without diagonal entries) as $G = (V_G, E_G)$ and build a group map ($\textit{gmap}: V_G \rightarrow \{1,\dots,|\textit{patches}|\}$) by lifting the mesh partition to the graph. In the scalar case, each mesh vertex maps to one graph vertex. In the block case, each mesh vertex maps to a small number of block of graph vertices, in which case we map all graph vertices in the block to the same patch. This covers both scalar Laplacian systems and block-structured Hessians. Figure~\ref{fig:Step1:Mapping} shows an instance of the scalar case. We use the resulting vertex-to-patch assignment to group the vertices of $G$. The map \textit{gmap} then assigns each vertex in $G$ to its corresponding group.

\subsection{Building the Elimination Tree}
\label{sec:etree}
Beyond producing a fill-reducing permutation, we construct and expose the \textit{etree}, which encodes Cholesky elimination dependencies and enables scheduling that trades locality for parallelism (\S\ref{sec:detail}). High-performance solvers (e.g., cuDSS) use the \textit{etree} to drive parallel numeric execution, and exposing it allows our permutation to integrate into such pipelines.

\begin{figure}
    \centering
    \includegraphics[width=0.9\columnwidth]{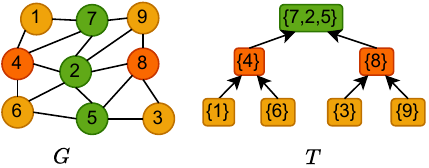}
    \caption{\textbf{Elimination tree construction.} We compute the root separator $\{7,2,5\}$. For the resulting subgraphs $\textit{g}_\ell=\{1,4,6\}$ and $\textit{g}_r=\{3,8,9\}$, we recurse to compute the next-level separators (e.g., $\{4\}$ and $\{8\}$), continuing until reaching the desired depth. We store the full binary tree in a 1D array.}
    \label{fig:Step2:BuildEtree}
\end{figure}

\paragraph{Step 3: Patch-guided nested dissection.}
Given the matrix graph $G$ and the group map $\textit{gmap}$, we recursively apply nested dissection for $\textit{nd\_level}$ levels while simultaneously constructing the \emph{etree} (Figure~\ref{fig:Step2:BuildEtree}). At each recursion, we compute a separator set $s$, partition the current subgraph into left and right subgraphs $g_\ell$ and $g_r$, store the separator at the current tree node, and recurse on the two children. The recursion terminates when the depth reaches zero, at which point the leaf stores the remaining vertices. We represent the \emph{etree} as a binary tree stored in a 1D array (node index $\textit{idx}$, children $2\textit{idx}+1$ and $2\textit{idx}+2$). In practice, moderate depths (e.g., $9$--$10$) provide sufficient parallelism~\cite{NVIDIA:2025:cudss}. While this procedure mirrors classical nested dissection, our runtime improves by targeting a dominant cost, i.e., separator computation at the top levels where subgraphs are largest and separators are most expensive as we discuss next.%

\begin{algorithm}[tp]
    \begin{algorithmic}[1]
        \Input $\textit{g}$, $\textit{gmap}$
        \Output $\textit{s}$, $\textit{g}_\ell$, $\textit{g}_r$
        \Statex \textcolor{gray}{/* 1) Build quotient graph over patch groups */}
        \State $\textit{q} \gets \textsc{QuotientGraph}(\textit{g}, \textit{gmap})$
        \Statex \textcolor{gray}{/* 2) Bipartition the quotient graph (with balance constraint) */}
        \State $\textit{part}_q \gets \textsc{BiPartition}(\textit{q})$
        \Statex \textcolor{gray}{/* 3) Lift partition and form a separator superset on $g$ */}
        \State $\textit{super\_s} \gets \textsc{SuperSeparator}(\textit{g}, \textit{gmap}, \textit{part}_q)$
        \Statex \textcolor{gray}{/* 4) Refine the superset (balance + local refinement) */}
        \State $\textit{s} \gets \textsc{RefineSeparator}(\textit{g}, \textit{super\_s}, \textit{gmap}, \textit{part}_q)$
        \Statex \textcolor{gray}{/* 5) Induce left/right subgraphs for the next recursion level */}
        \State $(\textit{g}_\ell, \textit{g}_r) \gets \textsc{LeftRight}(\textit{g}, \textit{s}, \textit{gmap}, \textit{part}_q)$
    \end{algorithmic}
    \caption{\textsc{GetSeparator} (called by \textsc{BuildEtree}).}
    \label{alg:getSeparator}
\end{algorithm}

\paragraph{Computing separators (\textsc{GetSeparator}).}
Our key idea is to reuse patch structure when extracting separators. Unlike classical multilevel nested dissection, we (i)~operate at the patch granularity using a quotient graph, rather than building a coarsening hierarchy, and (ii)~reuse the same patches across all recursion levels, trading some ordering quality for lower permutation overhead. 

Algorithm~\ref{alg:getSeparator} builds the quotient graph $q$ from the current patch group, bipartitions $q$, and lifts back to vertices in $V(g)$ to form a separator \emph{superset} of boundary vertices. We then refine this superset to reduce its size while maintaining balance, and extract the left and right subgraphs for recursion. Figure~\ref{fig:Step3:SeparatorComp} illustrates this pipeline.

\subsection{Permutation computation}
\label{sec:perm}

\paragraph{Step 4: Local orderings \& concatenation.}
Given the \emph{etree}, we compute the final permutation in two stages (Algorithm~\ref{alg:CompPerm}). At each tree node, we compute a \emph{local} ordering of its vertices (e.g., using AMD on the induced subgraph) and cache the result. Then, we concatenate these local orderings following a user-selected traversal of the \emph{etree} which we call the \emph{schedule}. \S\ref{sec:detail} explains the role of this order and the flexibility it provides for accelerating the numerical phase.

\begin{algorithm}[tp]
    \begin{algorithmic}[1]
        \Input $\textit{etree}$, $\textit{g}$, $\textit{schedule}$
        \Output $\textit{perm}$
        \Statex \textcolor{gray}{/* 1) Compute local permutations per tree node */}
        \For{$\textit{idx}$ in $[0, \ldots, |\textit{etree}|-1]$}
            \State $\textit{g}_{\textit{idx}} \gets \textsc{InducedSubgraph}(\textit{g},\, \textit{etree}[\textit{idx}].\textit{nodes})$
            \State $\textit{etree}[\textit{idx}].\textit{perm\_local} \gets \textsc{LocalOrder}(\textit{g}_{\textit{idx}})$
        \EndFor
        \Statex \textcolor{gray}{/* 2) Concatenate to form global permutation */}
        \State $\textit{perm} \gets [\,]$
        \For{$\textit{idx}$ in $\textit{schedule(etree)}$}
            \State $\textit{perm}.\textsc{Append}(\textit{etree}[\textit{idx}].\textit{perm\_local})$
        \EndFor
    \end{algorithmic}
    \caption{\textsc{ComputePerm}: local orderings and concatenation.}
    \label{alg:CompPerm}
\end{algorithm}



\section{Implementation Details}
\label{sec:detail}
If the user supplies a grouping of vertices (patches) such that each group induces a connected subgraph in the matrix graph, our method consumes it as the $\textit{gmap}$ and proceeds with Steps~3--4. If the provided groups are not connected, the algorithm still runs, but permutation quality can degrade since disconnected groups restrict \textsc{GetSeparator} and typically produce larger and less balanced separators. See Supplemental Materials B 
for more information about how the quality of patching correlates with separator size and runtime.

\begin{figure}
    \centering
    \fbox{\includegraphics[width=0.28\columnwidth]{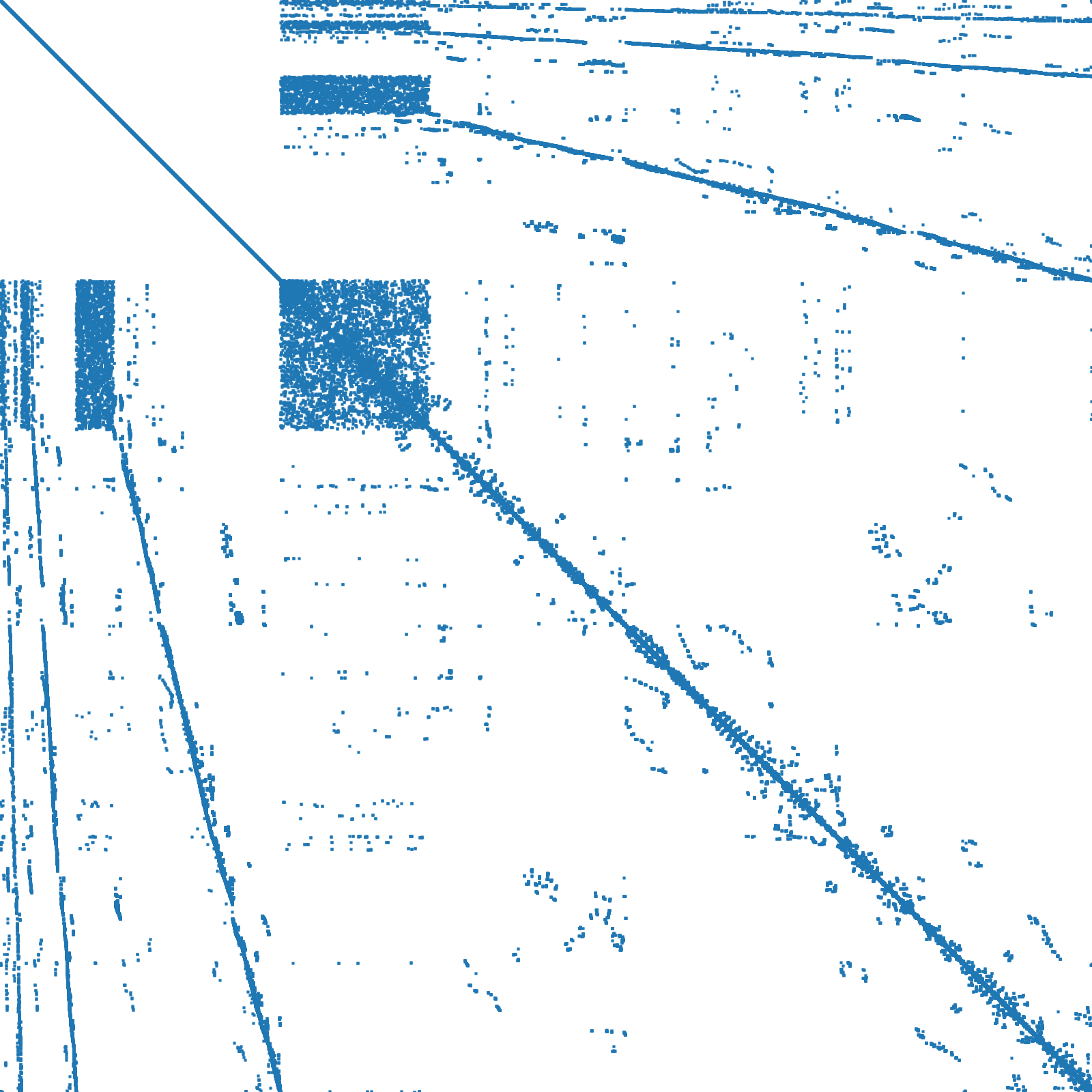}}    
    \fbox{\includegraphics[width=0.28\columnwidth]{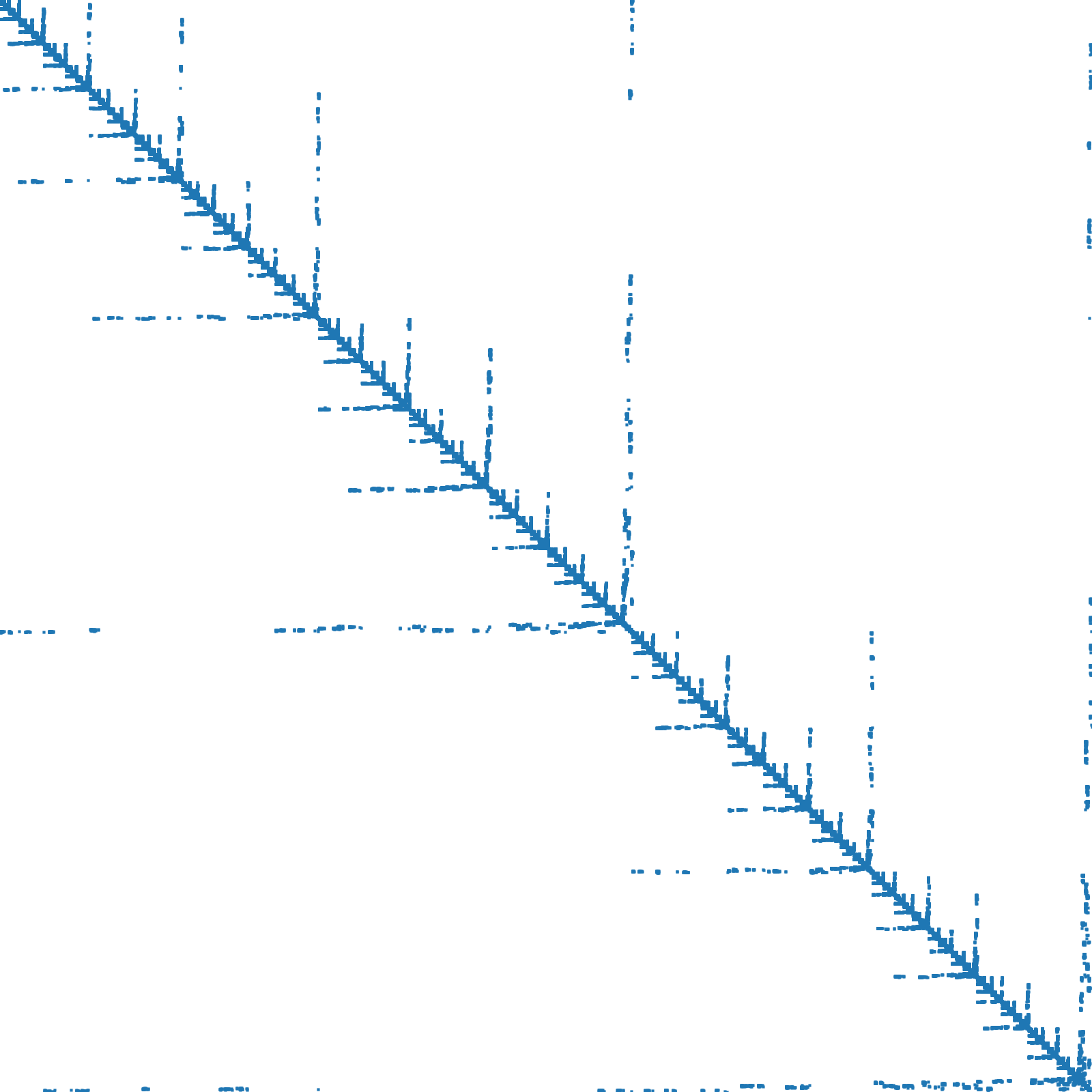}}
    \fbox{\includegraphics[width=0.28\columnwidth]{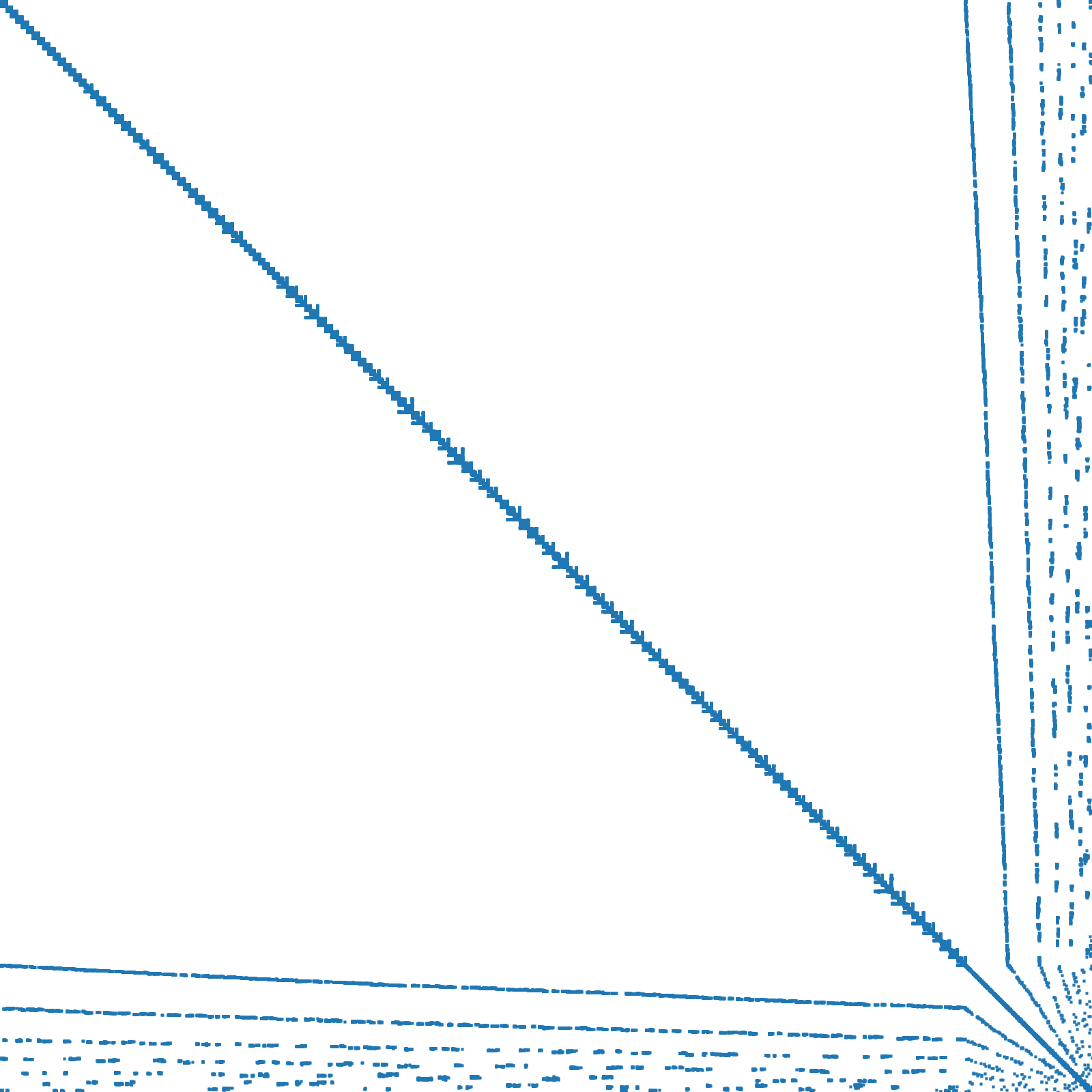}}
    \caption{For the same input (left), different schedules for traversing the elimination tree produce different permutations. Shown are post-order traversal (middle) and level-order traversal (right), which both result in the same fill-in.}
    \label{fig:schedule}
\end{figure}

\paragraph{Quotient graph computation.}
The quotient graph must be updated at each elimination tree node to produce a valid bipartition. However, a key cost we aim to avoid, especially for large matrices, is repeatedly scanning the full matrix. Since constructing the quotient graph from scratch typically requires reading the entire matrix, we instead build the quotient graph once and then update only node weights and edge weights as we compute \textit{etree} nodes from the root to the leaves. More specifically, we adjust the quotient graph node weights by removing the weight of the (already marked) separators, and adjust the edge weights by removing the edges of the separator nodes.

\paragraph{Separator set computation.} 
In separator set computation, we first use the quotient graph and METIS to find a bipartition. This step is fast because quotient graphs are relatively small compared to matrix $A$. Next, we obtain a superset of the separator by mapping the nodes in $V_G$ to the two partitions using their corresponding quotient nodes. This step is the bottleneck of our computation. Finally, we apply METIS's node refinement strategy to refine the separator. After finding the separator, we remove the separator nodes' effects from the quotient graph so that it can be reused in consecutive recursive calls of nested dissection. See Supplemental Materials B for more implementation details about separator refinement.

\paragraph{Concatenation and schedules.}
After computing the elimination tree, we form the global permutation by concatenating the local permutations associated with tree nodes (Algorithm~\ref{alg:CompPerm}). The primary degree of freedom is the \emph{schedule} used to traverse the elimination tree (Algorithm~\ref{alg:CompPerm}, line~6). A valid schedule must respect elimination dependencies, i.e., no parent can appear before its children. %
Different valid schedules do not change fill-in but can yield different performance due to locality--parallelism trade-offs. On CPUs, post-order traversal is preferred because it tends to improve temporal locality, i.e., once two sibling subtrees are processed, their results are consumed shortly thereafter by the parent, increasing cache reuse. This is the default in CPU solvers, e.g., CHOLMOD~\cite{Chen:2008:A8C}. In contrast, GPU solvers often prefer schedules that expose more parallelism, e.g., cuDSS follows a wavefront (level-based) schedule in which nodes are processed level by level starting from the leaves (see Figure~\ref{fig:schedule}).

\begin{figure}
\centering
    \includegraphics[width=0.98\columnwidth]{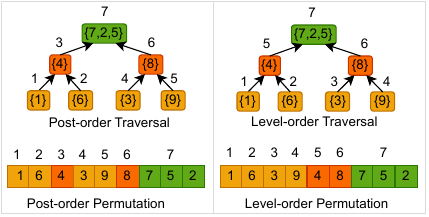}
\caption{Permutation vectors from post-order (left) and level-order (right) traversals of the \textit{etree}. Both yield the same fill-in, but can differ in performance due to memory access and parallelism.}
\label{fig:Step4:PermComp}
\end{figure}

Our implementation provides both post-order and level-order schedules (Figure~\ref{fig:Step4:PermComp}). More generally, the index sequence used in line~6 of Algorithm~\ref{alg:CompPerm} can be any valid schedule. Note that some CPU solvers (e.g., CHOLMOD) don't expose a user-defined schedule. Instead, they construct an internal execution order after \textit{etree} construction and then apply the given permutation. In such cases, providing an invalid schedule won't cause a runtime failure but can degrade performance if separators are assembled before their children. In contrast, cuDSS requires a schedule consistent with its execution model (i.e., a level-order/wavefront schedule); using an incompatible schedule can lead to runtime failure.

\section{Evaluation}
\label{sec:eval}
To evaluate the performance impact of our permutation algorithm and its portability across solvers and hardware platforms, we integrate it into two vendor-maintained sparse Cholesky solvers, i.e., NVIDIA cuDSS and Intel MKL. In both, we replace their fill-reducing permutation stage with ours. For cuDSS, we also replace the elimination-tree construction. Subsequent numerical factorization and solve routines are left unchanged, isolating the effect of our permutation on end-to-end performance.
We experiment on a system with an Intel Xeon Gold 6248 CPU (20 cores, 2.5 GHz, 28 MB LLC, 202 GB RAM) and an NVIDIA RTX 3080 GPU. We use Intel MKL version 2023.4-912 on Ubuntu 24.04 and CUDA 12.7 with cuDSS version 0.7.1. The input meshes we use in these experiments are collected from the Smithsonian~\cite{Smithsonian:2023:S3D}, ThreeDScans~\cite{Laric:2023:TDS}, and Thingi10K~\cite{Zhou:2016:TAD} repositories. We use double precision in all experiments.

Our evaluation proceeds in two parts. We first measure end-to-end solver speedups obtained by integrating our algorithm into cuDSS and Intel MKL across applications. We then study the permutation stage in isolation to identify the sources of these speedups. Together, these experiments capture common usage patterns of sparse Cholesky solvers in graphics applications. Unless otherwise stated, patches are computed on the GPU using Lloyd's algorithm~\cite{Lloyd:1982:LSQ} with a target patch size of 256 vertices. For comparison, we use METIS version shipped with cuDSS. We use the default parameters for cuDSS, i.e., double precision, complete GPU execution for factorization and solve, and default nested-dissection METIS.

\begin{table}[t]
  \centering
  \caption{\textbf{Application Speedup Summary.}
    We consider applications under (i) fixed and changing numerical factorization (F1: fixed, F2: changing), and (ii) different right-hand-side (RHS) use patterns (R1: repeated RHS solves, R2: matrix RHS). \emph{Mesh size} denotes the number of mesh vertices. \emph{\#iter} is the iteration at which we report the total solve speedup (e.g., for R1+F2 it counts refactorizations with the same sparsity). \emph{max \#iter} is the estimated break-even iteration count beyond which the default cuDSS ordering (METIS) becomes faster, if the lower-quality ordering reduces factorization and RHS-solve performance.}
  \label{tab:Evaluation:PerApp}

  {\small
  \begin{tabular}{p{0.24\columnwidth} l l c c c}
    \toprule
    App & Setting & Mesh size & \#iter & speedup & max \#iter \\
    \midrule

    \rowcolor{cyan!15}
    Data Smoothing & F1,R1 & 1M &  1 &   5.23x &      --- \\

    Data Smoothing & F1,R1 &      0.1M &      1 &      2.92x &    --- \\
    \midrule

    \rowcolor{cyan!15}
    SCP        & F1,R1    & 1.5M    & 32      & 4.16x    & 108859      \\

    SCP        & F1,R1    & 0.16M     & 32      & 2.34x    & inf         \\
    \midrule

    \rowcolor{cyan!15}
    Mesh Smoothing & F2,R2    & 1.76M    & 6       & 3.70x    & 160         \\

    Mesh Smoothing & F2,R2    & 0.17M     & 6       & 2.89x    & 148         \\
    \midrule

    \rowcolor{cyan!15}
    Parameterization  & F2,R1 & 1.1M &    13 &       1.47x &        53 \\

    Parameterization & F2,R1 & 0.32M  &    24 &       1.69x &       167 \\
    \bottomrule
  \end{tabular}}
\end{table}

\begin{figure}
    \centering
    \includegraphics[width=0.49\columnwidth]{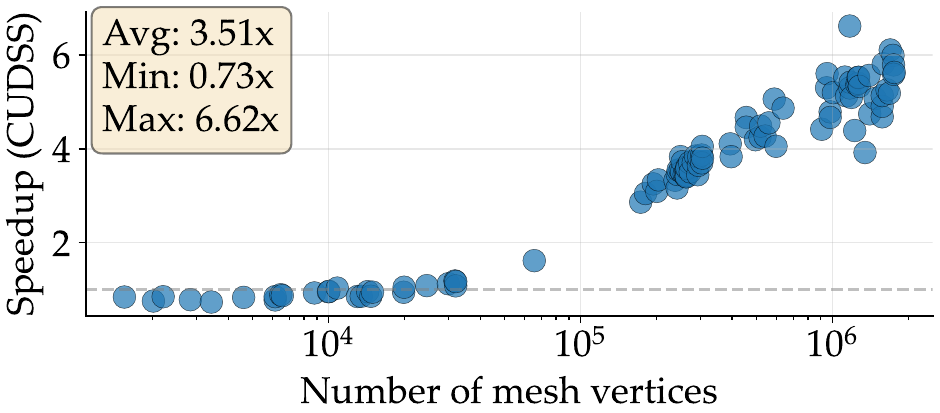}
    \includegraphics[width=0.49\columnwidth]{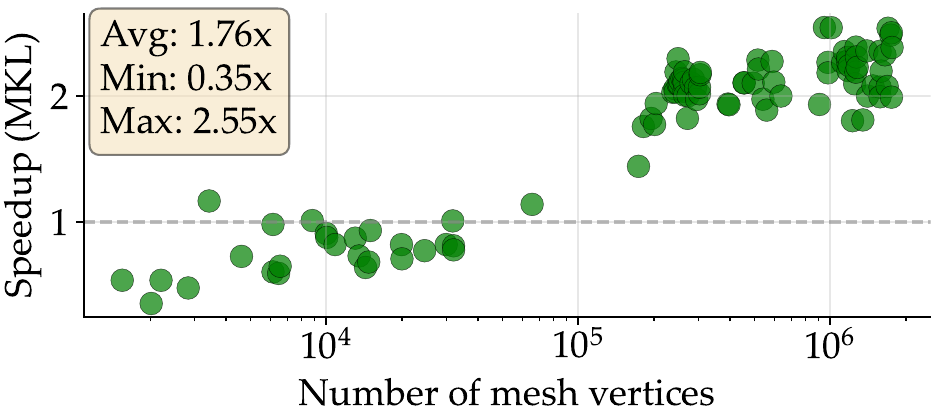}
    \caption{Speedup of cuDSS (left) and MKL (right) when replacing METIS with our permutation. Averaged across inputs, we achieve on average (geometric mean) $3.51\times$ on GPU and $1.76\times$ on CPU by reducing permutation overhead while preserving parallel-friendly elimination trees.}
    \label{fig:B1:TotalSpeedup}
\end{figure}
\subsection{Laplace-Beltrami}
\label{sec:laplace}
We begin with the Laplace-Beltrami operator assembled using libigl's \texttt{cotmatrix}~\cite{Jacobson:2018:LAS}. For each mesh, we perform a single sparse Cholesky factorization followed by a solve with a random right-hand side. This workload represents pipelines in which the system is factorized once and used immediately without amortizing symbolic preprocessing. In this regime, the symbolic phase can dominate runtime with fill-reducing permutation accounting for a large fraction of the cost (Figure~\ref{fig:perm_ratio}). To quantify our method's impact, we integrate it into cuDSS and Intel MKL and measure end-to-end speedups across 103 meshes with 1.5K to 1.8M vertices (Figure~\ref{fig:B1:TotalSpeedup}).

Replacing METIS with our algorithm yields increasing speedups with problem size. For large meshes, reduced permutation overhead directly yields faster solves as permutation increasingly dominates runtime. For meshes larger than 1.5M vertices, we observe speedups of up to $6.62\times$ with cuDSS and $2.55\times$ with Intel MKL. For smaller meshes, the benefits are limited where METIS can perform better since the permutation cost is small (Figure~\ref{fig:perm_ratio}).
Empirically, our method becomes beneficial above approximately 50K vertices on cuDSS and 100K vertices on Intel MKL. %
In Supplemental Materials D, 
we compare against the recently introduced multi-threaded mode in cuDSS---which was released concurrently with our work. In addition, we study the case of refactorizations (i.e., same sparsity with different numerical values) and report the break-even iteration after which our method becomes slower in Supplemental Materials E.

For the remaining evaluation, we focus exclusively on cuDSS since our patching stage is GPU-accelerated, so pairing it with a GPU solver provides a natural end-to-end view. In addition, cuDSS inputs both a permutation and an (\emph{etree}), allowing us to evaluate permutation overhead as well as the quality of the \emph{etree} and its impact on factorization and solve performance.

\subsection{Data Smoothing}
\label{sec:data_smoothing}
\begin{figure}
    \centering
    \includegraphics[width=0.49\textwidth]{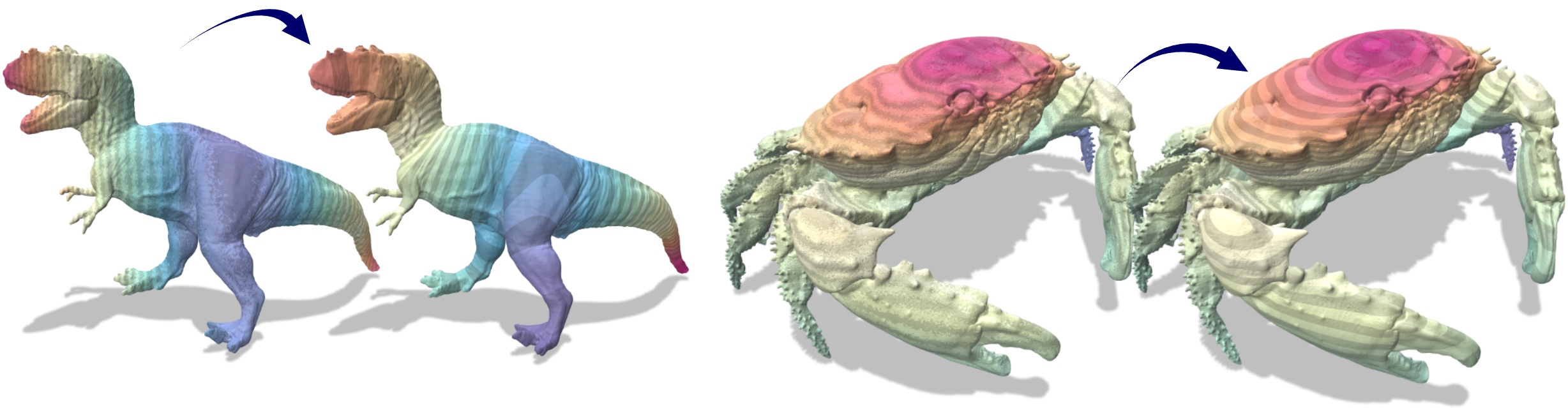}
    \caption{\textbf{Data smoothing via Laplacian regularization.} We recover a smooth scalar function from noisy vertex data by solving a quadratic optimization problem that balances a mass-weighted data term with a Laplacian-based smoothness term.}
    \label{fig:data_smooth}
\end{figure}

Data smoothing on triangle mesh recovers a smooth scalar function $x \in \mathbb{R}^{|V_M|}$ from a noisy signal $y$ defined on the mesh vertices (Figure~\ref{fig:data_smooth}). We compute $x$ by minimizing a quadratic energy that balances data fidelity and smoothness using Dirichlet energy: $\min_x (x - y)^\top M (x - y) + \alpha x^\top L^\top M^{-1} L x$, where $M$ is the lumped mass matrix, $L$ is the cotangent Laplacian, and $\alpha > 0$ controls the smoothing. Setting the gradient to zero yields $\big( (1-\alpha) M + \alpha \, L^\top M^{-1} L \big) x = \alpha M y$, which we solve using Cholesky factorization.

Table~\ref{tab:Evaluation:PerApp} reports end-to-end speedups for a large mesh (1M vertices) and a smaller mesh (100K vertices). We achieve $5.23\times$ and $2.92\times$ speedup on the 1M-vertex mesh and 100K-vertex mesh, respectively.  This mirrors the trend in the Laplace-Beltrami evaluation (\S\ref{sec:laplace}), since both involve a single sparse Cholesky factorization where permutation cost dominates at larger sizes.

\begin{figure}
    \centering
    \includegraphics[width=0.49\textwidth]{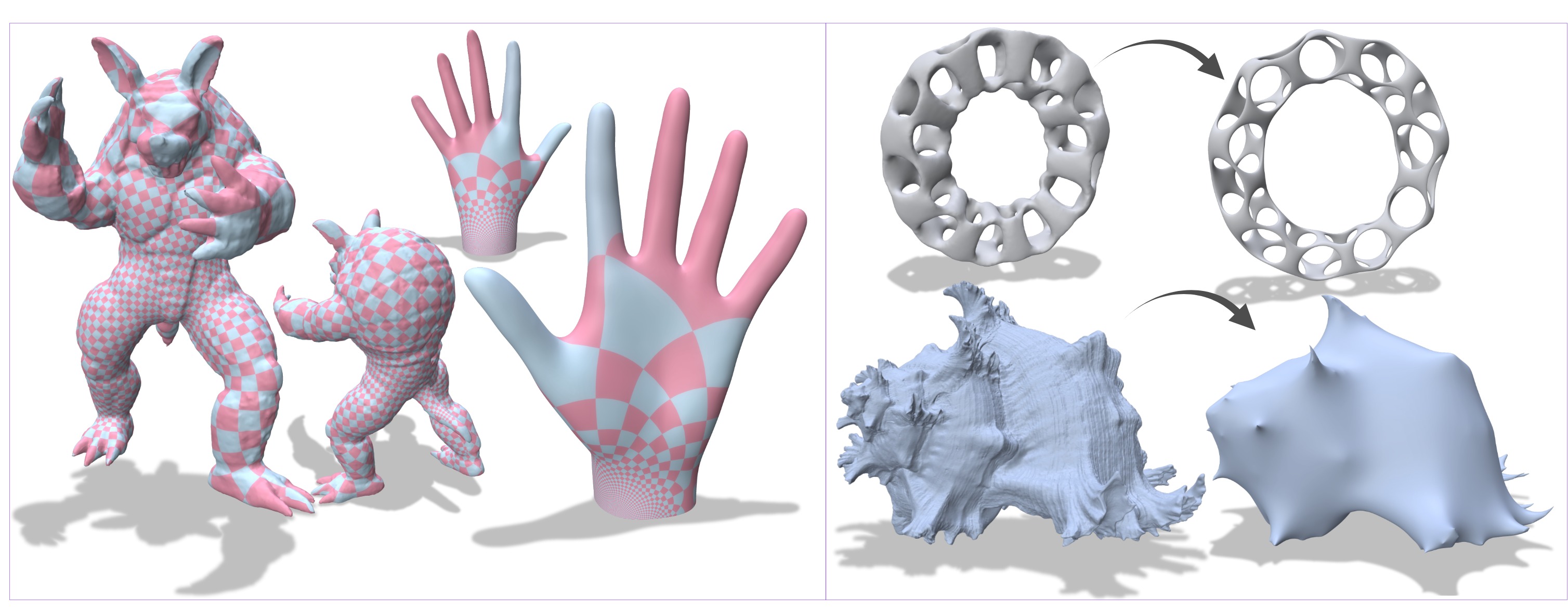}
    \caption{Results of SCP produced by our implementation (left) and results of six iterations of mesh smoothing applied on a mesh with 1.7M vertices produced by our implementation (right).}
    \label{fig:scp_and_smoothing}
\end{figure}

\subsection{Spectral Conformal Parameterization}
\label{sec:scp}

We next evaluate on spectral conformal parameterization (SCP)~\cite{Mullen:2008:SCP}, a parameterization method that solves a generalized eigenvalue problem derived from the discrete conformal energy (see Figure~\ref{fig:scp_and_smoothing}). In practice, SCP is implemented using inverse power iteration, where each iteration solves a sparse linear system with a fixed matrix and different right-hand side. In our implementation, the conformal energy matrix and its sparsity pattern remain fixed. The solver performs a single sparse Cholesky factorization followed by multiple forward/backward substitutions. SCP represents a workload where symbolic analysis and factorization are amortized while permutation quality affect the cost of repeated solves.

We integrate our permutation into cuDSS and measure end-to-end speedup using 32 power iterations. Table~\ref{tab:Evaluation:PerApp} shows that replacing METIS leads to $4.16\times$ and $2.34\times$ on the larger and small mesh, respectively. Since forward and backward substitution incur much lower overhead than factorization, accelerating permutation and symbolic preprocessing has a lasting impact when many solves are performed. For smaller meshes, our permutation can improve solve performance relative to METIS. For larger meshes, the patch-based approximation and local AMD permutation may slightly degrade factor quality and increase substitution cost. This effect is not systematic and depends on the sparsity pattern. We analyze this trade-off in \S\ref{sec:benchmark}. Overall, SCP demonstrates a complementary regime to the Laplace-Beltrami experiment where a single factorization is reused across many solves. Reducing permutation overhead still yields end-to-end gains even when factorization is amortized.

\subsection{Mesh Smoothing}
\label{sec:mesh_smoothing}
\begin{figure}
    \centering
    \includegraphics[width=0.48\textwidth]{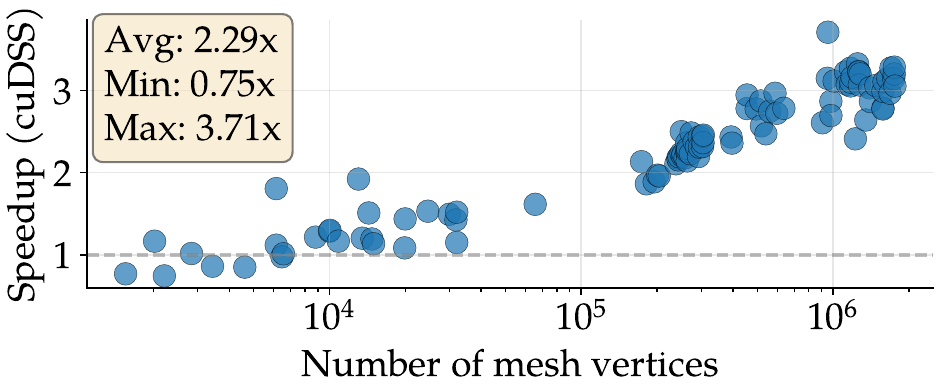}
    \caption{\textbf{End-to-end solver speedup on mesh smoothing.} Using cuDSS, our fill-reducing permutation achieves an average $2.29\times$ speedup across six smoothing iterations, with larger meshes benefiting the most.}
    \label{fig:smoothing_benchmark}
\end{figure}
We next consider iterative smoothing via mean-curvature flow~\cite{Desbrun:1999:IFO} (see Figure~\ref{fig:scp_and_smoothing}). Each iteration assembles a new system matrix, followed by sparse Cholesky (re)~factorization and a solve. The right-hand side is a dense $|V_M|\times 3$ matrix which we solve simultaneously. This workload combines repeated factorizations with multiple right-hand-side solves, exhibiting a different balance between permutation, factorization, and substitution costs.

After six iterations, our permutation yields a $3.7\times$ speedup for the large mesh and $2.89\times$ for the small mesh (Table~\ref{tab:Evaluation:PerApp}). The break-even point (measured in repeated full numerical phases) is 160 iterations for the large mesh and 148 for the small mesh. Although our permutation sometimes can lead to slightly slower factorization and triangular solves, overall performance still favors our pipeline due to the substantial reduction in the main computational bottleneck, i.e., permutation. Figure~\ref{fig:smoothing_benchmark} shows the end-to-end speedup over cuDSS across different mesh sizes. As seen before, speedup increases with problem size, reaching up to $3.71\times$ for the largest meshes.

\begin{figure*}
    \centering
    \includegraphics[width=0.32\textwidth]{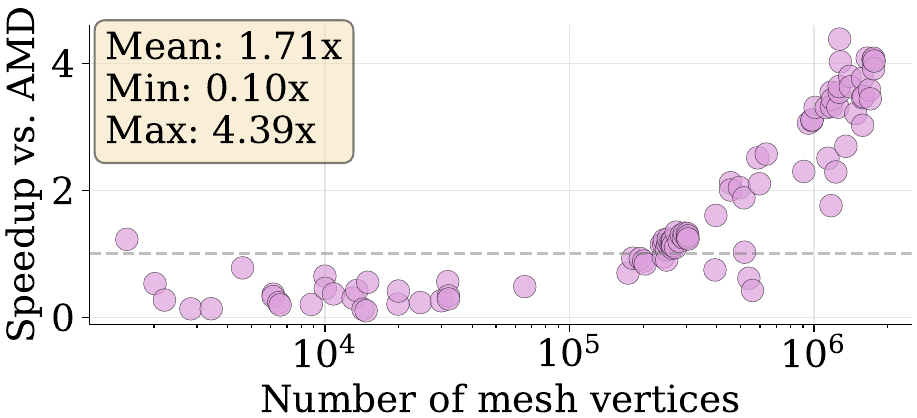}\includegraphics[width=0.32\textwidth]{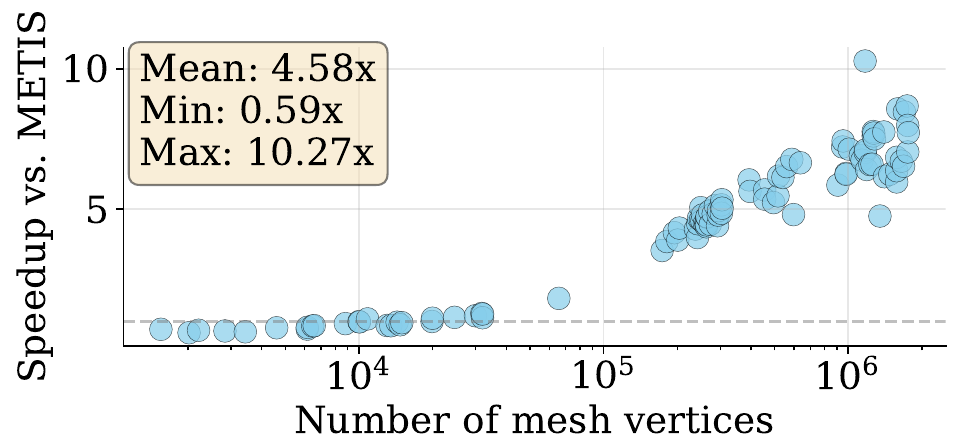}\includegraphics[width=0.32\textwidth]{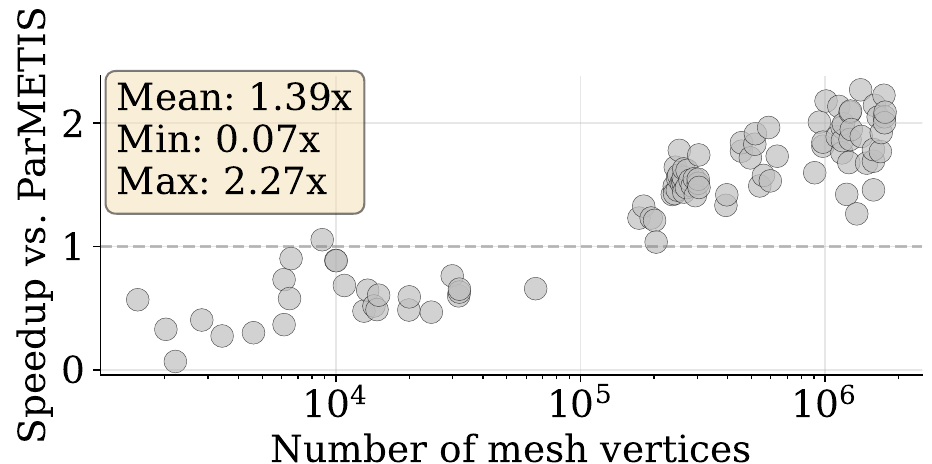}
    \caption{\textbf{Runtime comparison of our algorithm vs.\ AMD, METIS, and ParMETIS.} Our algorithm scales better than the baseline ordering tools commonly integrated into high-performance linear solvers, and begins to outperform them on meshes with more than 100k vertices.}
    \label{fig:PermRuntimeComp}
\end{figure*}

\subsection{Symmetric Dirichlet Parameterization}
\label{sec:sym_dirichlet}
We minimize the symmetric Dirichlet energy~\cite{Smith:2015:BPW} to compute a low-distortion 2D parameterization of a surface mesh. We solve the resulting nonlinear system using Newton's method where each iteration assembles and factorizes a sparse Hessian and then solves for a Newton step. This application has similar characteristic to mesh smoothing (\S\ref{sec:mesh_smoothing}), where matrix numerical values change while the sparsity is constant. Unlike the Laplacian-based workloads in previous sections, this application produces a second-order system. The Hessian sparsity pattern is closely related to the Laplace-Beltrami operator but differs in scale, as each nonzero in the Laplacian corresponds to a dense $2\times2$ Hessian block. Consequently, the Hessian has $2\times$ as many rows/columns and $\sim4\times$ as many nonzeros as the scalar Laplacian.

This increased density makes factorization more expensive, reducing the fraction of end-to-end time spent in permutation. Since sparse Cholesky has complexity $O\left(\sum_{i=1}^{N} d(i)^2\right)$~\cite{Lipton:1979:GND}, where $d(i)$ is the number of nonzeros in column $i$ of the factor $L$, replacing scalar entries with $2\times2$ blocks increases the effective column densities by roughly 2. For the 1M-vertex mesh, Laplacian factorization takes 72~ms, while Hessian factorization takes 300~ms. Hence, speedups here are smaller than for mesh smoothing (\S\ref{sec:mesh_smoothing}), permutation accounts for a larger share of runtime (see Table~\ref{tab:Evaluation:PerApp}).

Our permutation cost remains comparable to the Laplacian case since we apply it to a compressed graph where we merge each $2\times2$ block into a single node when constructing the matrix graph $G$. We expand the resulting permutation and \emph{etree} back to the full Hessian system.%

\subsection{Benchmark and Ablation}
\label{sec:benchmark}

We benchmark along two axes: (i)~permutation runtime and (ii)~fill-in quality. We first compare the permutation time against AMD, METIS, and ParMETIS (Intel oneAPI) to isolate the cost of permutation independent of factorization. We then analyze our algorithm in more detail. First, we study \emph{patch-size sensitivity} by varying the patch size and measuring fill-in. Second, we present a \emph{module runtime breakdown}, reporting the cost of each stage, including patch construction, patch-based separator computation, and local AMD refinement. Finally, we study \emph{design choices in patch ordering} by comparing runtime and fill-in when (i)~replacing patch-based separators with METIS separators and (ii)~using AMD vs.\ METIS for local permutation. These experiments justify our default configuration. We additionally report the speedup of all applications using Intel MKL as a backend in Supplemental Materials F.

\paragraph{Baseline Comparison}
Figure~\ref{fig:PermRuntimeComp} shows ordering speedups relative to AMD, METIS, and ParMETIS. While our method is less competitive on small problems ($n<10^5$), it scales substantially better with mesh size by restricting separator search to a smaller quotient graph and reusing patch structure across recursion levels. For $n>10^5$, this yields speedups of up to $10.27\times$ over METIS and $2.27\times$ over ParMETIS.

\begin{figure}
    \centering
    \includegraphics[width=0.45\textwidth]{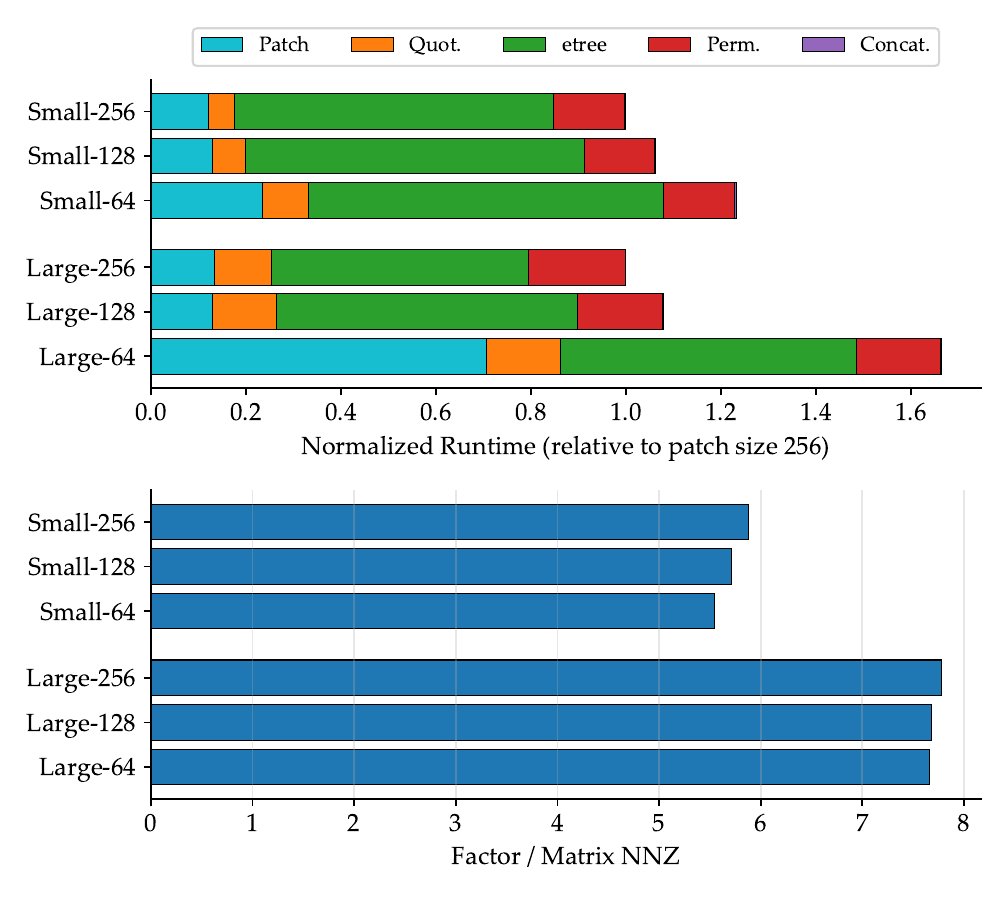}
    \caption{\textbf{Patch size trade-off in our algorithm.}
        Top: runtime (lower is better) breakdown of each stage (patching, patch-based separator computation, quotient-graph/\emph{etree} construction, and local AMD ordering) for patch sizes ${64,128,256}$, shown for a small ($n \approx 300$K) and a large mesh ($n \approx 1.8$M). Each stacked bar is normalized by the total ordering time at patch size 256. Bottom: permutation quality measured by the fill-in ratio $\mathrm{nnz}(L)/\mathrm{nnz}(A)$ (lower is better). Larger patches reduce permutation time by producing fewer patches and a smaller quotient graph but increase fill-in by restricting separator flexibility. For patch size 256, the total ordering time of our algorithm is 1.717s (large) and 0.207s (small), compared to 10.272s and 1.418s for METIS, respectively.}
    \label{fig:PatchSizeComp}
\end{figure}
\paragraph{Patch Size Effect}
Figure~\ref{fig:PatchSizeComp} shows the trade-off induced by patch size. We vary the target patch size in $\{64,128,256\}$ and report (i)~runtime breakdowns of each stage in our algorithm and (ii)~permutation quality measured by the nonzero ratio $\mathrm{nnz}(L)/\mathrm{nnz}(A)$.

\begin{figure}
    \centering
    \includegraphics[width=0.98\columnwidth]{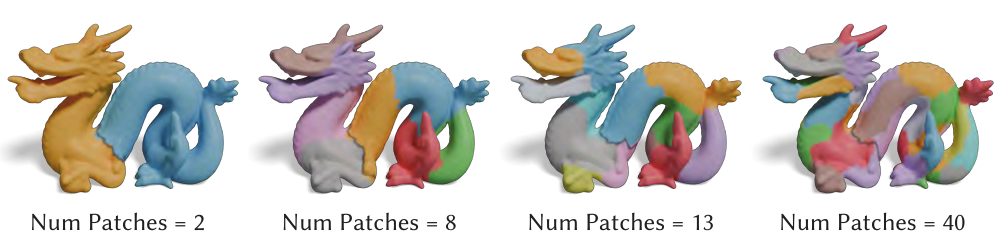}
    \caption{\textbf{Effect of patch size on separator quality.} Decreasing the patch size results in a larger number of patches, allowing our algorithm to produce higher-quality separators, since the separator must follow patch boundaries. However, this comes with a trade-off where increasing the number of patches also increases runtime overhead. %
    }
    \label{fig:PatchSizeEffect}
\end{figure}
Larger patches reduce permutation time by producing fewer patches and a smaller quotient graph, lowering both patch-processing overhead and \emph{etree} construction cost. This is visible in the top plot, where patch sizes 128 and 256 reduce runtime relative to patch size 64, especially on the large mesh. The downside is reduced separator flexibility which can degrade permutation quality and increase fill-in. This is reflected in the bottom plot, where larger patches yield slightly higher $\mathrm{nnz}(L)/\mathrm{nnz}(A)$ (see Figure~\ref{fig:PatchSizeEffect}). We discuss in more details the runtime breakdown of our permutation algorithm in Supplemental Materials G.

The trade-off is most pronounced on the smaller mesh where coarse patches over-constrain separator placement and increase fill-in. In contrast, the large mesh is less sensitive to patch size, as more patches preserve sufficient flexibility. In this regime, a patch size of 256 provides a favorable balance and we use it as the default.

\paragraph{Patch-based Separator Computation}
\begin{figure}
    \centering
    \includegraphics[width=0.45\textwidth]{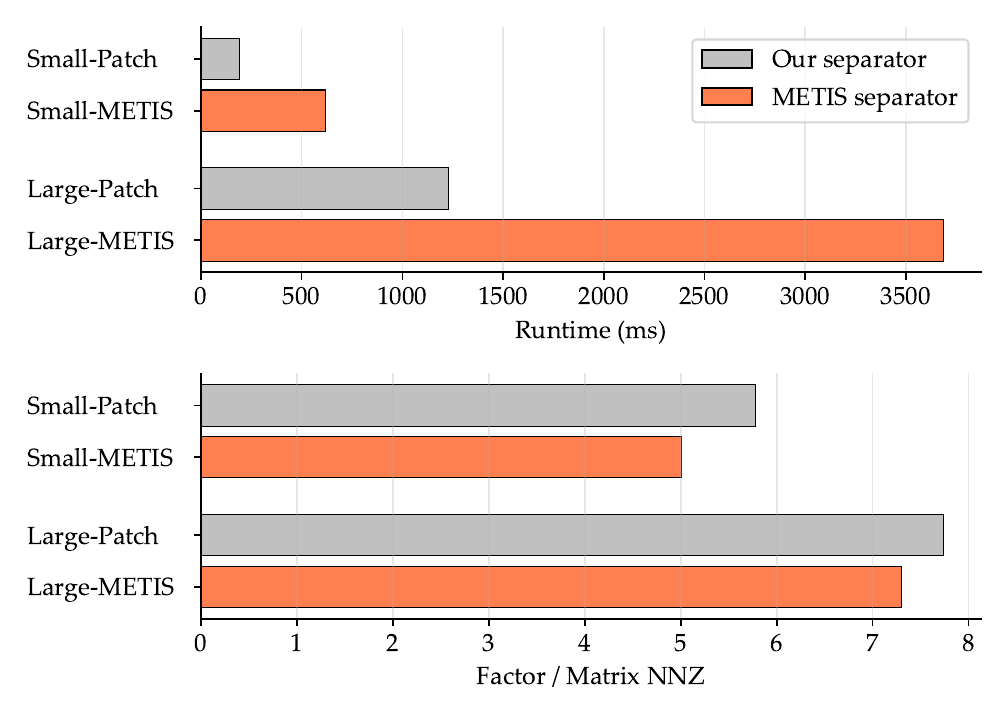}
    \caption{\textbf{Separator computation and separator refinement: METIS vs.\ our patch-based algorithm.} We compare METIS vertex separator computation and refinement against our algorithm's patch-based separator stage at patch size 256, on a small mesh and a large mesh. \texttt{Small-Patch} and \texttt{Small-METIS} denote the corresponding method/mesh pair. Top: the patch-based runtime (lower is better) includes patch construction, quotient-graph construction, and \emph{etree} construction (cf.\ Figure~\ref{fig:PatchSizeComp}). Bottom: permutation quality measured by the fill-in ratio $\mathrm{nnz}(L)/\mathrm{nnz}(A)$ (lower is better).}
    \label{fig:SeparatorComp}
\end{figure}

A key design choice is to compute separators on the quotient graph rather than on the full matrix graph, reducing separator cost by shrinking the search space and reusing patch structure across recursion levels. We compare our patch-based separator stage (excluding local refinement) against \texttt{METIS\_Compute\-Vertex\-Separator}~\cite{Karypis:2013:MSG} (Figure~\ref{fig:SeparatorComp}). On the small and large meshes, patch-based computation is $3.22\times$ and $3\times$ faster, respectively, with $15.44\%$ and $6.07\%$ increases in $\mathrm{nnz}(L)/\mathrm{nnz}(A)$. These results show that quotient-graph separators substantially reduce runtime with modest fill-in increases and illustrate how patch size trades off runtime against ordering quality.

\paragraph{Local Permutation Effect}
\begin{figure}
    \centering
    \includegraphics[width=0.45\textwidth]{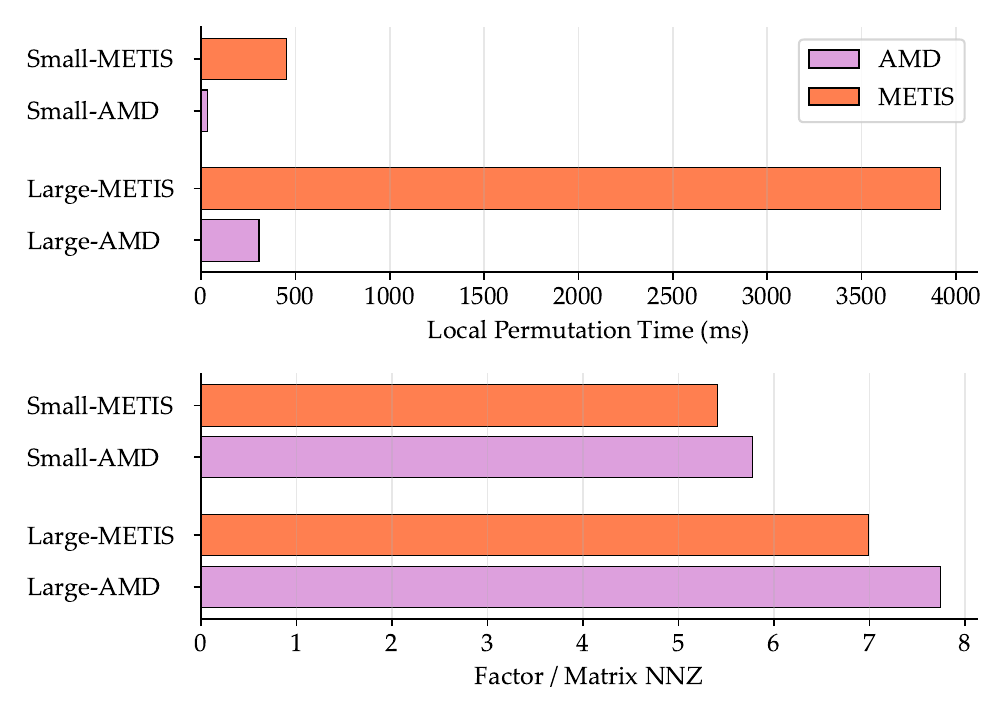}
    \caption{\textbf{Local permutation trade-off: AMD vs. METIS.} Local permutation time (top, lower is better) and fill-in ratio $\mathrm{nnz}(L)/\mathrm{nnz}(A)$ (bottom, lower is better) for AMD and METIS used as the local permutation step within \emph{etree} construction in our algorithm with patch size 256. AMD yields an order-of-magnitude reduction in local ordering time, at the cost of a modest increase in fill-in.}
    \label{fig:ModularEffect}
\end{figure}

Finally, we examine the local permutation used within each \emph{etree} subgraph. We compare AMD (default) against METIS at patch size 256. Figure~\ref{fig:ModularEffect} shows that AMD substantially reduces local-permutation time. On the small and large meshes, AMD is $12.42\times$ and $12.74\times$ faster, respectively, with $6.67\%$ and $10.78\%$ increases in $\mathrm{nnz}(L)/\mathrm{nnz}(A)$. Overall, AMD provides an order-of-magnitude reduction in local ordering time at the cost of a modest increase in fill-in motivating our default choice when minimizing permutation overhead is the primary objective. 

We expand on this ablation in Supplemental Materials H 
and examine the effect of separator and local permutation on the factorization and triangular-solve runtime.



\section{Conclusion, Limitations, and Future Works}
\label{sec:conclusion}
Sparse Cholesky is the method of choice for solving SPD systems due to its robustness but it scales poorly. We target the main bottleneck in this pipeline, i.e., fill-reducing permutation. We design a simple algorithm for fast matrix permutation with a focus on end-to-end runtime. In doing so, we trade some permutation quality for speed. Our results show this is worthwhile for many use cases, including single solves, repeated solves, and block-structured Hessians.

This trade-off is beneficial when the solve phase is executed only a small number of times before refactorization or a change in matrix structure. In such settings, our method yields a net speedup (\S\ref{sec:eval}). However, when the matrix remains fixed and the solve phase is repeated for thousands of iterations, it can cause a small end-to-end slowdown. For example, in inverse rendering, the same factorized matrix is reused to solve many right-hand sides (see Supplemental Materials I 
for details and results).

We achieve this by specializing our algorithm to systems from triangle meshes, where domain structure provides strong guidance for permutation. Extending to other domains is promising, with tetrahedral meshes as a natural next step. In a preliminary tet-mesh experiment using \texttt{METIS\_KWay} for patch generation, our method achieves up to $3.73\times$ speedup without patching, but drops to $0.83\times$ when the patching time is included. This indicates that volumetric meshes will require a fast tetrahedral patching routine. Future work also includes GPU-based separator computation and parallel local ordering.



\begin{acks}
\label{sec:ack}
The authors would like to thank Jonathan Ragan-Kelley for valuable discussions. The MIT Geometric Data Processing Group acknowledges the generous support of National Science Foundation grants IIS2335492 and OAC2403239, from the CSAIL Future of Data and FinTechAI programs, from the MIT--IBM Watson AI Laboratory, from the Wistron Corporation, from the MIT Generative AI Impact Consortium, from the Toyota--CSAIL Joint Research Center, and from Schmidt Sciences.
\end{acks}


\bibliographystyle{ACM-Reference-Format}




\includepdf[pages=-]{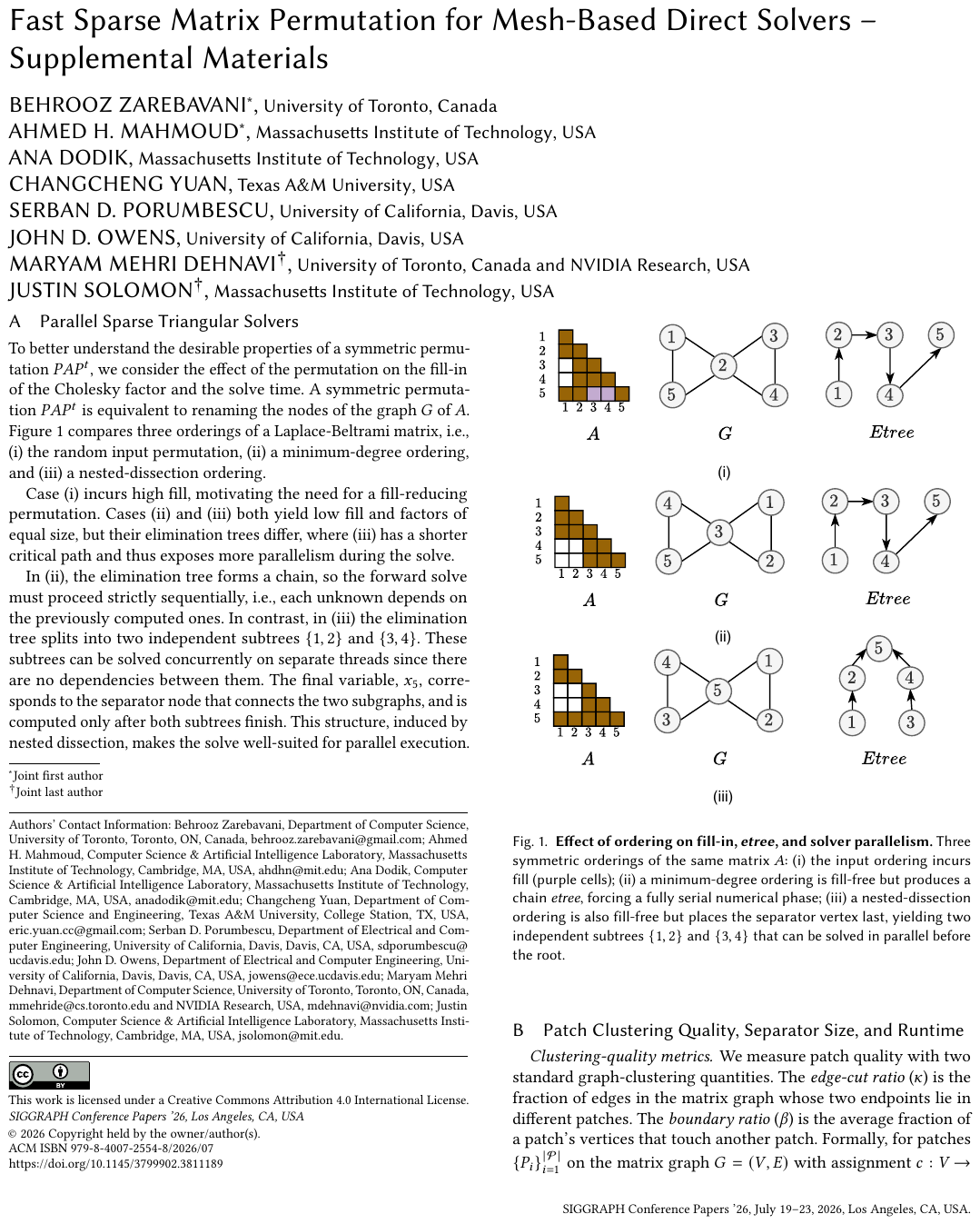}
\end{document}